\documentclass[]{pasj01}
\Received{$\langle$reception date$\rangle$}
\Accepted{$\langle$acception date$\rangle$}
\Published{$\langle$publication date$\rangle$}

\begin{document}

\title{Possible Identification Of Massive and Evolved Galaxies At $z \gtrsim 5$}
\author{K. \textsc{Mawatari}\altaffilmark{1}, T. Yamada\altaffilmark{2}, G. G. Fazio\altaffilmark{3}, J. -S. Huang\altaffilmark{3}, and M. L. N. Ashby\altaffilmark{3}}%
\altaffiltext{1}{College of General Education, Osaka Sangyo University, 3-1-1, nakagaito, Daito, Osaka 574-8530, Japan}
\altaffiltext{2}{Institute of Space Astronautical Science, Japan Aerospace Exploration Agency, Sagamihara, Kanagawa 252-5210, Japan}
\altaffiltext{3}{Harvard-Smithsonian Center for Astrophysics, 60 Garden St., Cambridge, MA 02138, USA}
\email{mawatari@las.osaka-sandai.ac.jp}

\KeyWords{galaxies: evolution --- galaxies: formation --- galaxies: high-redshift}

\maketitle

\begin{abstract}
We report on the identification of the old stellar population galaxy candidates at $z \gtrsim 5$. We developed a new infrared color selection scheme to isolate galaxies with the strong Balmer breaks at $z \gtrsim 5$, and applied it to the ultra-deep and wide infrared survey data from the $Spitzer$ Extended Deep Survey (SEDS) and the UKIRT Infrared Deep Sky Survey. The eight objects satisfying $K - [3.6] > 1.3$ and $K - [3.6] > 2.4 ([3.6] - [4.5]) + 0.6$ are selected in the 0.34 deg$^2$ SEDS Ultra Deep Survey field. Rich multi-wavelength imaging data from optical to far-infrared are also used to reject blending sources and strong nebular line emitters, and we finally obtained the three most likely evolved galaxies at $z \gtrsim 5$. Their stacked spectral energy distribution is fitted well with the old stellar population template with M$_{*} = (7.5\pm1.5) \times 10^{10}$\,M$_{\odot}$, star formation rate $= 0.9 \pm 0.2$\,M$_\odot$\,yr$^{-1}$, dust $A_V < 1$, and age $= 0.7\pm0.4$\,Gyr at $z = 5.7\pm0.6$, where the dusty star-forming galaxies at $z \sim 2.8$ is disfavored because of the faintness in the $24\,\mu$m. The stellar mass density of these evolved galaxy candidates, $(6 \pm 4) \times 10^4$\,M$_{\odot}$\,Mpc$^{-3}$, is much lower than that of star-forming galaxies, but the non-zero fraction suggests that initial star-formation and quenching have been completed by $z \sim 6$. 
\end{abstract}

\section{Introduction}\label{sec:intro}
\indent

When and how did massive quiescent or evolved galaxies like present-day elliptical galaxies form? According to the current $\Lambda$ cold dark matter paradigm, dark matter halos grow by assembling material from their neighborhood, or by merging with other halos. This view is supported by cosmological simulations \citep{WhiteRees78,Davis85,Springel05,Mo10}. On the other hand, the growth of galaxies, the baryonic components of the halos, seems not to proceed in a simple bottom-up manner. Observations have shown that many of the most massive galaxies (M$_{star}$ $>$ 10$^{11}$M$_{\odot}$) in the local universe assembled quickly and at early epochs ($z > 2$) while less massive galaxies grew more gradually \citep{Kodama98,Marchesini09,Marchesini10,Ilbert10,Brammer11,Muzzin13}. It is clear that the baryonic physics of galaxy formation involves more than just cooling and accretion of gas at a rate dictated by gravity. 

In order to understand the balance among several processes in galaxy formation [major/minor mergers, harassment, feedback by active galactic nuclei (AGNs) or supernovae, cold gas accretion, and so on], it is important to unravel the relationship among different galaxy populations such as actively star forming galaxies, quiescent galaxies, and AGNs. The first quiescent galaxies to emerge are especially interesting because their redshifts and masses will have much to say about the history of galaxy formation and star formation activity in the very early universe. However, our present knowledge of galaxies at these epochs are based primarily on star-forming galaxies and AGNs, due to the limited sensitivity and volume of surveys carried out to-date, as well as an intrinsic rarity of quiescent galaxies at high redshift \citep{Kajisawa11,
Dominguez11,Muzzin13}. A new, larger sample of quiescent galaxies at higher redshift is needed to obtain a more complete picture of galaxy formation and evolution at early epochs. 

In contrast with star-forming galaxies, which typically exhibit numerous emission lines or strong Lyman breaks, quiescent galaxies are relatively difficult to be identified at high redshift because of their nearly featureless spectral energy distributions (SEDs). The Balmer/4000-\AA-break is the only strong feature in the SEDs of quiescent galaxies, and it has long been used to identify them at $z = 1 - 3$ \citep{Elston88,Dickinson95,Yamada97,Yamada05,Franx03,Hatch11}. For example, objects with $R - K > 5$ or $I - K > 4$ in the Vega magnitude system select elliptical galaxies with a strong 4000\AA-break at $z \sim 1$ (Extremely Red Objects: EROs; \cite{Thompson99,MacCarthy01,Cimatti02,Caputi04}). This ERO selection was extended to higher redshift, $z \sim 2.4$, by adopting the red $z - [3.6]$ color criterion ($f_{\nu}[3.6\mu{\rm m})/f_{\nu}(z) > 20$, IRAC-selected extremely red object (IERO); \cite{Yan04}]. The near-infrared (NIR) color criterion $J - K > 2.3$ (Vega) is often used to select quiescent galaxies at $z = 2 - 3$ (Distant Red Galaxies: DRGs; \cite{vanDokkum03,Kajisawa06b,Kubo13}). While dusty star-forming galaxies also satisfy these simple $R - K$ and $J - K$ color criteria, a substantial fraction of EROs and DRGs are spectroscopically confirmed to be passively evolved galaxies \citep{Cimatti02,Cimatti04,Doherty05,Kriek06a,Kriek06b}. As spectroscopy is time-consuming, many authors have instead separated passive and dusty star-forming galaxies photometrically using two colors so that one color brackets the Balmer break and the other color represents the shape of spectrum at longer wavelengths. The $JHK$ color selection [$J - K > 2(H - K) + 0.5$ and $J - K > 1.5$ in the Vega magnitude system] is similar to the DRG selection, but more sensitive to the evolved galaxy population at $z \sim 2.5$ \citep{Kajisawa06a,Doherty10}. It is also known that passive and dust-obscured galaxies at $z > 2$ separate well in the $I - K$ vs. $K - [4.5]$ two color diagram \citep{Labbe05,Papovich06,Kubo13}. Recently, rest-frame $U - V$ and $V - J$ (rest-$UVJ$) color selection is favored, which allows us to distinguish passive galaxies from star-forming galaxies independently of their redshifts as long as SED fitting works well \citep{Wuyts07,Williams09,Whitaker11,Brammer11,Morishita14}.

At $z > 3$, the Balmer/4000\AA\ break is shifted to $\lambda > 1.6\mu$m, and detecting quiescent galaxies at those redshifts is a challenging task. However, some candidates at $z = 4 - 6$ have been found in extremely deep NIR and IR surveys \citep{Rodighiero07,Wiklind08, Mancini09,Huang11,Caputi12,Nayyeri14}. \citet{Huang11} analyzed very deep {\it Hubble Space Telescope} ($HST$) and $Spitzer$ imaging in the Great Observatories Origins Deep Survey South (GOODS-South) field and identified four galaxies with $H - [3.6] > 4.5$. Huang et al. (2011) argued that the four extremely red objects are likely distant passive galaxies rather than nearby dusty galaxies, because their Multiband Imaging Photometer (MIPS) 24-$\mu$m fluxes are weaker than expected if their rest-infrared SEDs are similar to those of the central region of M82 (a local dusty galaxy with $5 < A_V < 51$) or those of the dust obscured galaxies (DOGs) at $z \sim 2$. Employing a similar strategy, \citet{Caputi12} identified 25 objects with $H - [4.5] > 4$ over $\sim$ 180 arcmin$^2$ of the $HST$ Cosmic Assembly Near-infrared Deep Extragalactic Legacy Survey (CANDELS; \cite{Grogin11,Koekemoer11}) Ultra Deep Survey (UDS) field. Their sample includes one $z \sim 6$ massive galaxy candidate whose SED is fitted well with a 0.5-Gyr-old stellar population template with a stellar mass of $3 \times 10^{11}$\,M$_{\odot}$. Many other red $H - [4.5]$ objects also prefer high-redshift photometric solutions ($3 < z < 5$), but their red SEDs are consistent with significant dust extinction: $1.1 \le A_{V} \le 4.2$ \citep{Caputi12}. \citet{Wiklind08} proposed a color selection that is more sensitive to galaxies with strong Balmer breaks (Balmer Break Galaxies: BBGs) at $z = 4.9 - 6.5$, and which is based on the three colors, $K - [3.6]$, $J - K$, and $H - [3.6]$. They adopted their selection to the multi-band imaging available in the GOODS-South field, and identified the 11 candidates of passive galaxies with ages $0.2 - 1$\,Gyr at $z \gtrsim 5$. They mentioned that the number of candidates decrease to 4 if they allow only objects undetected in the MIPS 24$\mu$m.

In this work, we applied a similar color selection of high-redshift evolved galaxies to the 3.6$\mu$m and 4.5$\mu$m imaging from the $Spitzer$ Extended Deep Survey (SEDS: \cite{Ashby13}) and the $K$-band imaging from the United Kingdom Infrared Telescope (UKIRT) Infrared Deep Sky Survey (UKIDSS: \cite{Lawrence07}). The $K - [3.6]$ color is expected to be more sensitive to the Balmer break at $z=5 - 6$ than $H - [3.6]$ or $H - [4.5]$ because it spans a relatively short wavelength interval. The wide area of the UKIDSS UDS (\cite{Almaini??}) enables us to search for rare galaxies over the entire SEDS UDS field (0.34\,deg$^2$), whereas the survey area of the previous studies using $HST$ are relatively small. The aim of this work is to identify evolved or old stellar population galaxies at $z \gtrsim 5$.

This paper is structured as follows. In Section 2 we summarize data used in this work. We describe the $K - [3.6]$ versus $[3.6] - [4.5]$ two-color selection scheme and further selection using multi-wavelength imaging in Section 3. Stacking analysis and results for the most likely evolved galaxy candidates are described in Section 4. In Section 5, we discuss the implications about the high-$z$ evolved galaxies. Our conclusions are summarized in Section 6. Unless specified otherwise, we use the AB magnitude system \citep{OkeGunn83} and adopt a cosmology with $H_{0}=70.4$ km s$^{-1}$ Mpc$^{-1}$, $\Omega_{M}=0.272$, and $\Omega_{\Lambda}=0.728$ \citep{komatsu11}.

\section{Data}\label{sec:data}

\begin{figure}[t]
\begin{center}
\includegraphics[width=1.0\linewidth, angle=0]{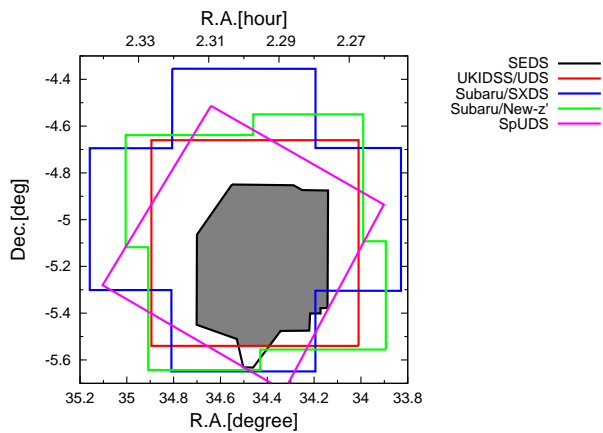}
\end{center}
\caption{Coverages of the survey data used in this study: SEDS, UKIDSS, SXDS, New z'-band, and SpUDS images. HerMES covers all other FoVs. We investigated objects in the overlapped region of SEDS and UKIDSS (grey shaded region). \label{fig1}}
\end{figure}

\begin{table*}
\caption{UDS DATASET\label{tb1}}
\begin{center}
\begin{tabular}{llcccl}
\hline
\hline
Instrument & Filter & FWHM\footnotemark[$*$] & Area & Limiting magnitude\footnotemark[$\dagger$] & Survey\footnotemark[$\ddagger$] \\
 &  & (arcsec) & (deg$^2$) & (5$\sigma$, AB) &  \\
\hline
Subaru/Suprime-Cam & $B$ & 0.80 & 1.23 & 27.5 & (1)\\
Subaru/Suprime-Cam & $V$ & 0.80 & 1.23 & 27.1 & (1)\\
Subaru/Suprime-Cam & $R$ & 0.80 & 1.23 & 26.9 & (1)\\
Subaru/Suprime-Cam & $i'$ & 0.80 & 1.23 & 26.8 & (1)\\
Subaru/Suprime-Cam & $z'$ & 1.0 & 1.00 & 26.3 & (2)\\
UKIRT/WFCAM & $J$ & 0.79 & 0.78 & 24.9 & (3)\\
UKIRT/WFCAM & $H$ & 0.80 & 0.78 & 24.3 & (3)\\
UKIRT/WFCAM & $K$ & 0.75 & 0.78 & 24.5 & (3)\\
$Spitzer$/IRAC & 3.6$\mu$m & 1.8 & 0.34 & 25.4 & (4)\\
$Spitzer$/IRAC & 4.5$\mu$m & 1.8 & 0.34 & 25.4 & (4)\\
$Spitzer$/IRAC & 5.8$\mu$m & 2.1 & 0.82 & 21.7 & (5)\\
$Spitzer$/IRAC & 8.0$\mu$m & 2.3 & 0.80 & 21.5 & (5)\\
$Spitzer$/MIPS & 24$\mu$m & 5.9 & 1.06 & 19.2 & (5)\\
$Herschel$/SPIRE & 250$\mu$m & 18.2  & 2.0 & 13.5 & (6)\\
$Herschel$/SPIRE & 350$\mu$m & 24.9  & 2.0 & 13.5 & (6)\\
$Herschel$/SPIRE & 500$\mu$m & 36.3  & 2.0 & 13.3 & (6)\\
\hline
\end{tabular}
\end{center}
\begin{tabnote}
\footnotemark[$*$] FWHMs were estimated by measuring the mean values of those of 100 - 200 bright stars in the SEDS area except for HerMES. The PSFs for HerMES are quoted from Oliver et al. (2012).\\
\footnotemark[$\dagger$] Limiting magnitudes were measured adopting 2 $\times$ PSF (FWHM) diameter aperture except for SEDS and HerMES data. Limiting magnitudes of SEDS are quoted from \citet{Ashby13}. For the HerMES imaging, the limiting magnitudes are estimated by using the HerMES error maps \citep{Levenson10}.\\
\footnotemark[$\ddagger$] (1) SXDS - \citet{Furusawa08} (2) New data - \citet{Furusawa16} (3) UKIDSS UDS DR10 - \citet{Lawrence07,Almaini??} (4) SEDS - \citet{Ashby13} (5) SpUDS - P.I.: Dunlop (6) HerMES - \citet{Oliver12,Wang14}
\end{tabnote}
\end{table*}

\indent 

This research is primarily based on data from SEDS \citep{Ashby13}, a very deep infrared survey using $Spitzer$'s Infrared Array Camera (IRAC: \cite{Fazio04}) to cover five well-known extragalactic science fields. SEDS reaches a depth of 25.4 mag (0.25\,$\mu$Jy; 5$\sigma$) in both of the warm IRAC bands at 3.6 and 4.5 $\mu$m. In this work we analyze the $\sim$ 0.34 deg$^2$ of the SEDS UDS field, where rich multi-wavelength imaging is available from a diverse array of ground- and space-based facilities. The surveys used in this study are summarized in Table \ref{tb1} and Figure \ref{fig1}. 

The visible-wavelength photometry comes from the Subaru/XMM-Newton Deep Survey (SXDS) which covers the entire SEDS UDS area, reaching 5$\sigma$ limits (2$''$ diameter aperture) of $B = 27.5$, $V = 27.1$, $R = 26.9$, $i' = 26.8$ \citep{Furusawa08}. For $z'$-band imaging, we used the new data obtained in 2008 - 2009 (P.I.: H. Furusawa; \cite{Furusawa16}) with Suprime-Cam \citep{Miyazaki02} on the Subaru Telescope. This new $z'$-band imaging reaches the 5$\sigma$ limits of $z' = 26.3$ (2$''$ diameter aperture), which is $\sim$ 0.5 mag deeper than that of the SXDS $z'$-band imaging.

The depth and coverage of the near-infrared imaging are the most important factors in this study. Deep $K$- or $H$-band imaging is required to detect or constrain the flux at shorter wavelengths than the Balmer break of quiescent galaxies at $z \gtrsim 5$. We obtained the deep $J$-, $H$-, and $K$-band imaging from the UKIDSS UDS (DR10$^{1}$), which have $\sim$ 0.8 deg$^2$ coverage containing nearly all of the SEDS UDS coverage. The UKIDSS project is defined in \citet{Lawrence07}. Further details on the UDS can be found in \citet{Almaini??}. UKIDSS uses the UKIRT Wide Field Camera (WFCAM; \cite{Casali07}). The photometric system is described in \citet{Hewett06}, and the calibration is described in \citet{Hodgkin09}. The pipeline processing and science archive are described in Irwin et al. (in prep) and \citet{Hambly08}.  In principle, wide area coverage should increase the detection rate for quiescent galaxies at high redshift because their spatial distribution is expected to be highly biased, which is the confirmed trend at $1 < z < 3$ in the previous studies of EROs or DRGs \citep{Daddi00,Daddi03,Roche02,Hamana06,Tinker10,Uchimoto12}. The UDS reaches depths of $J = 24.9$, $H = 24.3$, and $K = 24.5$ (5$\sigma$). Compared with the CANDELS/$J$ and $H$-band imaging, the UDS $K$-band imaging yields the advantage in the wide area and small wavelength interval between $K$ and 3.6$\mu$m. We estimated the above 5$\sigma$ limiting magnitudes of the UDS by measuring the standard deviation of the background fluctuations with 2$''$ diameter aperture. Our estimated limiting magnitudes are $\sim$ 0.3 - 0.4 mag shallower than the official values reported by the WFCAM Science Archive\footnote{http://surveys.roe.ac.uk/wsa/dr10plus\_release.html}. For the present work, we adopt our own measurements as the conservative limiting magnitudes. 


While no survey achieved the same depth as SEDS in the mid-infrared range, shallower IRAC/5.8, 8.0, and MIPS/24$\mu$m imaging from the $Spitzer$ UKIDSS Ultra Deep Survey (SpUDS; P.I.: J. Dunlop) may be useful for constraining the SEDs of galaxies. SpUDS is shallow, but covers a wide area ($\sim 1$\,deg$^2$), which contains the entire SEDS UDS coverage. We obtained the SpUDS imaging data through NASA/IPAC Infrared Science Archive (IRSA), which is operated by the Jet Propulsion Laboratory, California Institute of Technology, under contract with the National Aeronautics and Space Administration. We found that the limiting magnitudes are 21.7, 21.5, and 19.2 mag (7.6, 9.1, and 76\,$\mu$Jy; 5$\sigma$) in the 5.8, 8.0, and 24$\mu$m bands by measuring the standard deviations of the background fluctuations, where we adopted 2 $\times$ Full Width at Half Maximum (FWHM) of the Point Spread Function (PSF) diameter apertures (4.2, 4.6, and 12 arcsec). 

We also used SPIRE 250, 350, and 500$\mu$m imaging from the Herschel Multi-tiered Extragalactic Survey (HerMES: \cite{Oliver12}). In the second HerMES data release (DR2), SPIRE imaging in the L4-UDS field \citep{Oliver12} is available, which contains the entire SEDS UDS coverage. The HerMES data was accessed through the Herschel Database in Marseille (HeDaM)\footnote{http://hedam.lam.fr} operated by CeSAM and hosted by the Laboratoire d'Astrophysique de Marseille. The depth of the SPIRE imaging is estimated by using the corresponding error maps \citep{Levenson10}, finding that the limiting magnitudes are 14.7, 14.2, and 17.1\,mJy ($5\sigma$) in the 250, 350, 500$\mu$m bands. We used the HerMES band-merged source catalog (xID catalog: \cite{Wang14}) for the 250, 350, and 500 $\mu$m photometry.

\section{Candidate selection}\label{sec:selection}
\subsection{\it Multi-band photometry for SEDS-detected objects}\label{sec:multi-phot}
\indent

\begin{figure}[]
\begin{center}
\includegraphics[width=1.0\linewidth, angle=0]{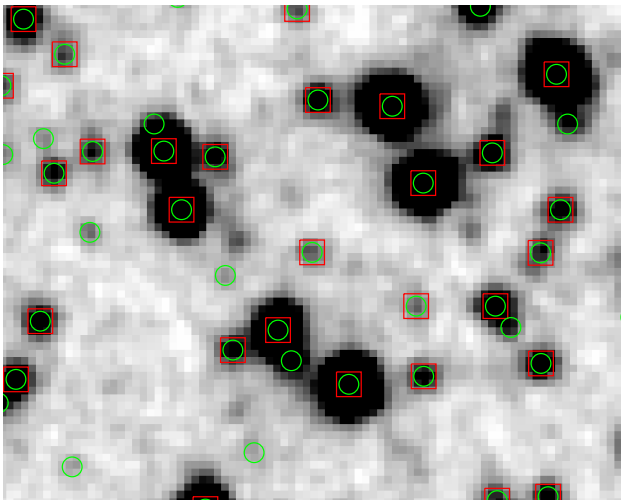}
\end{center}
\caption{Postage-stamp image of the SEDS 3.6$\mu$m, where the panel size is $50'' \times 40''$. Sources used in our analysis are shown by red squares having sides of 2$''$. Objects in the SEDS catalog \citep{Ashby13} are also shown by  green  circles. \label{fig2}}
\end{figure}

\begin{figure}[]
\begin{center}
\includegraphics[width=1.0\linewidth, angle=0]{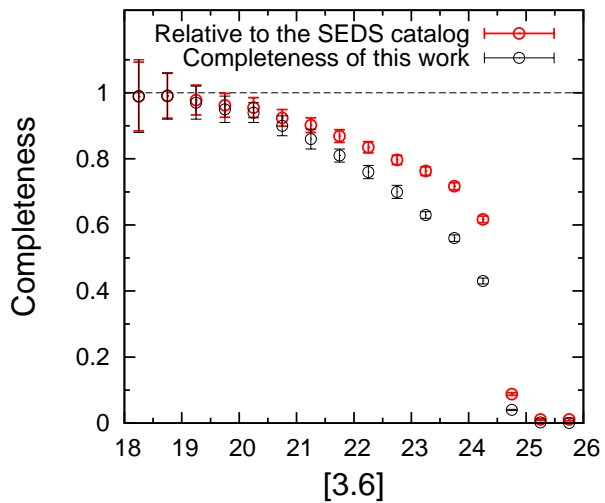}
\end{center}
\caption{Completeness of our source extraction in the 3.6$\mu$m image relative to that of the SEDS catalog (red circles). We also plotted the completeness that was estimated by multiplying that of the SEDS catalog \citep{Ashby13} and the relative completeness (black circles). \label{fig3}}
\end{figure}

We used the SEDS catalog \citep{Ashby13} for source extraction and photometry in the 3.6 and 4.5$\mu$m. The SEDS catalog yields PSF-fitted total magnitudes, from which we estimated 2.4\,arcsec aperture magnitudes using the aperture correction factors of \citet{Ashby13}. This small aperture size compared with the PSF (FWHM $= 1.8$\,arcsec) is recommended by the SEDS survey team. We also investigated signal to noise ratio ($S/N$) in the 3.6um photometry as a function of aperture size, the results of which were that $S/N$ decreases with increasing aperture size and the 2.4\,arcsec aperture is better. In this study, we focus on relatively isolated sources among the cataloged objects, because source confusion makes it hard to compare or combine multi-band photometry. In order to select isolated sources, we performed source extraction again on the 3.6$\mu$m image by using SExtractor \citep{BertinArnouts96} version 2.5.0. It is suggested that SExtractor does not resolve heavy source blending at faint flux level in IRAC images \citep{Sanders07,Ashby13}. The software was configured carefully so that we could extract as faint isolated sources as possible. We show the SExtractor parameter settings in Table \ref{tb2}. The objects detected in both of the SEDS catalog and our own SExtractor catalog are selected for our analysis. The objects used for our analysis are shown in Figure~\ref{fig2}, where all objects in the SEDS catalog are also shown for comparison. We failed to detect not only confused but also very faint sources, which does not change the result of this paper because such faint objects in 3.6$\mu$m cannot be identified as red $K - [3.6]$ objects, which we focus in the following sections. The completeness of our source extraction in the 3.6$\mu$m image relative to the SEDS catalog is shown in Figure~\ref{fig3}. By multiplying this relative completeness and the completeness of the SEDS catalog \citep{Ashby13}, we estimated the completeness of this work (Figure~\ref{fig3}).

\begin{table}[]
\caption{SExtractor configuration parameters for object extraction on the SEDS 3.6$\mu$m image.\footnotemark[$*$]\label{tb2}}
\begin{center}
\begin{tabular}{ll}
\hline\hline
Parameter & Value \\
\hline
DETECT\_MINAREA & 1.5\footnotemark[$\dagger$] \\
DETECT\_THRESH & 1.2\footnotemark[$\dagger$] \\
DEBLEND\_NTHRESH & 64 \\ 
DEBLEND\_MINCONT & 0.0001 \\
\hline
\end{tabular}
\end{center}
\begin{tabnote}
\footnotemark[$*$]{We set the default values for other parameters about object extraction.}\\
\footnotemark[$\dagger$] {Pixel unit.}
\end{tabnote}
\end{table}

We measured the 2.4$''$ aperture magnitudes using the photometry task ``PHOT'' of Image Reduction and Analysis Facility (IRAF) for the $B$, $V$, $R$, $i'$, $z'$, $J$, $H$, $K$, 5.8$\mu$m, and 8.0$\mu$m images, where each aperture was centered on the position of the object detected in the 3.6 $\mu$m image. The $B$-, $V$-, $R$-, $i'$-, $z'$-, $J$-, $H$-, and $K$-band images were convolved to the same resolution as that of the 3.6 $\mu$m and 4.5 $\mu$m images (FWHM $\sim$ $1\farcs8$). We corrected the $B$, $V$, $R$, $i'$, $z'$, $J$, $H$, and $K$ magnitudes for the Galactic extinction with A$_B = 0.091$, A$_V = 0.07$, A$_R = 0.056$, A$_{i'} = 0.044$, A$_{z'} = 0.031$, A$_J = 0.019$, A$_H = 0.012$, and A$_K = 0.0075$. These values were estimated for the position (RA, Dec: \timeform{02h18m00s}, \timeform{-05D00'00''} in J2000) in the UDS field based on the work by \citet{Schlegel98}, assuming R$_V = $A$_V/$E$(B-V) = 3.1$. When we measured the $2\farcs4$ diameter aperture magnitudes in the 5.8 and 8.0$\mu$m images, we corrected the difference of PSFs to be matched with those of the 3.6 and 4.5$\mu$m images assuming the same surface brightness profile in all bands (``negative aperture correction''). For the purpose, we smoothed the 4.5$\mu$m image to the same resolution as that of the 5.8 and 8.0$\mu$m images (FWHM $\sim$ $2\farcs1$ and $2\farcs3$), respectively. The correction factors were then obtained from the flux ratio of the original and smoothed images for each individual object. The average correction factor of the 5.8 and 8.0$\mu$m images are $-0.09$ and $-0.15$ mag, respectively. 

We checked whether our objects are sufficiently bright to be detected in the SpUDS 24$\mu$m and HerMES imaging. For the SpUDS 24$\mu$m image, object detection and photometry with 12$"$ diameter apertures was performed using SExtractor. For the HerMES images, the 250, 350, and 500$\mu$m fluxes were obtained for the objects in the HerMES DR1 catalog \citep{Oliver12,Wang14}.

\subsection{\it Color Selection of BBGs}\label{sec:CC}

\begin{figure*}[t]
\begin{center}
\includegraphics[width=1.0\linewidth, angle=0]{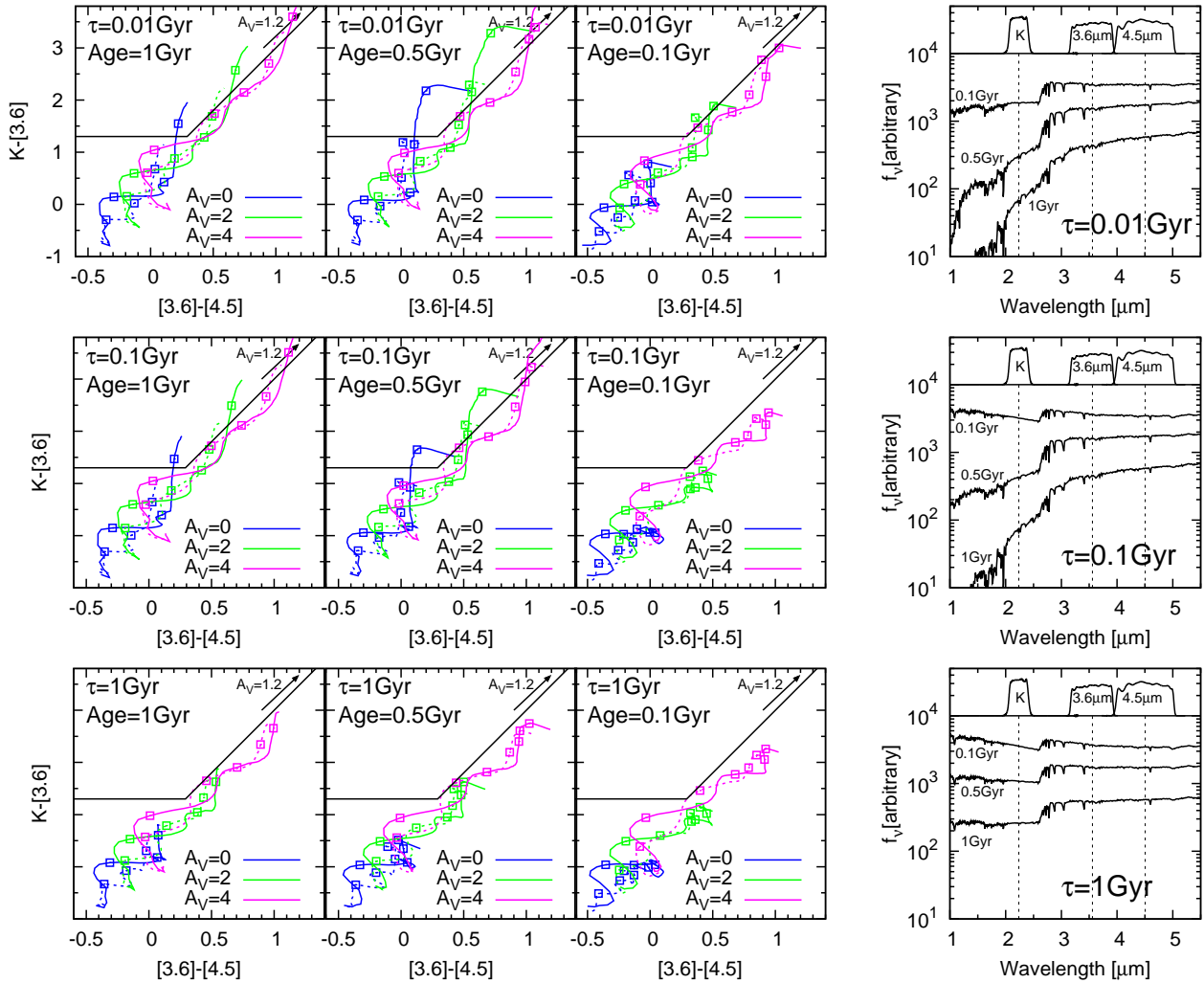}
\end{center}
\caption{Behaviors of various spectral templates in the $K - [3.6]$ versus $[3.6] - [4.5]$ two color diagram. The templates come from the BC03 stellar population synthesis models with the exponentially declining star-formation history. Star-formation timescale is set to $\tau = 0.01, 0.1$, and 1\,Gyr, and age is set to 0.1, 0.5, and 1\,Gyr. In each panel, we show the models with different A$_V$ (blue: A$_V = 0$; green: A$_V = 2$; magenta: A$_V = 4$) and different metallicity (solid: $Z = 0.02$; dashed: $Z = 0.0001$). The open squares superposed on each model track correspond to $z = 1, 3, 5, and 7$. The horizontal and diagonal black lines correspond to our BBG selection boundaries, where the diagonal boundary line is parallel to the reddening vector at $z = 5.3$ (black arrow). In the extreme right-hand side panels, we show the model template spectra with different star-formation timescale $\tau$ and age at $z = 6$, where the A$_V$ and metallicity are fixed (A$_V = 0$ and $Z = 0.02$). \label{fig4}}
\end{figure*}

\begin{figure*}[]
\begin{center}
\includegraphics[width=1.0\linewidth, angle=0]{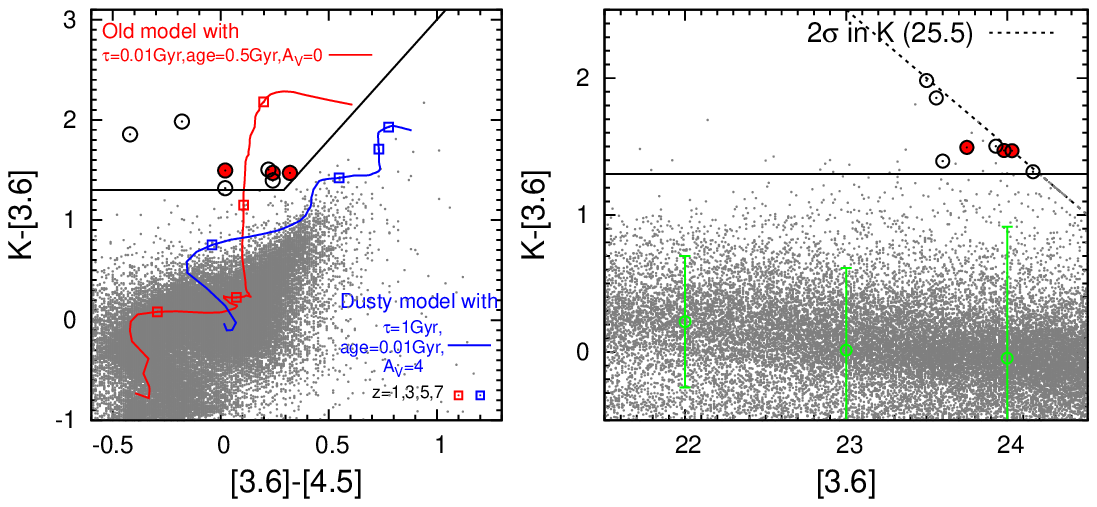}
\end{center}
\caption{(Left-hand panel) All objects detected in the SEDS image (grey points) in the $K - [3.6]$ vs. $[3.6] - [4.5]$ two color diagram. Black open circles are the BBG candidates selected by our color criteria (black horizontal and diagonal lines). Red filled circles are the most likely BBGs at $z \gtrsim 5$. Solid red and blue curves are the BC03 model tracks with different star-formation timescale, age, and A$_V$, where metallicity is fixed to $Z = 0.02$. (Right-hand panel) All objects (grey points), color-selected BBG candidates (black open circles), and most likely BBGs at $z \gtrsim 5$ (red filled circles) in the [3.6] versus $K - [3.6]$ color-magnitude diagram. The green circles show the peaks of $K - [3.6]$ color distribution as a function of 3.6$\mu$m magnitude, and error bars correspond to the 4$\sigma$ color errors estimated from the photometric uncertainties in the $K$ and 3.6$\mu$m images.  \label{fig5}}
\end{figure*}

\indent

Our sample selection of evolved galaxies with strong Balmer breaks (BBGs; \cite{Wiklind08}) is based on the $K - [3.6]$ and $[3.6] - [4.5]$ colors. We examined the colors of various types of galaxies using the stellar synthesis models of GALAXEV (\cite{BruzualCharlot03}, hereafter BC03). In Figure~\ref{fig4}, we show the behavior of the colors of the BC03 model galaxies over a wide variety of redshift, star formation history (SFH), age, metallicity, and dust extinction, where initial mass function (IMF) and dust extinction law were fixed at a Chabrier IMF \citep{Chabrier03} and Calzetti law \citep{Calzetti00}, respectively. It is notable that galaxies at $z \gtrsim 5$ with ages much older than their star-formation timescale ($\tau$) are red in $K - [3.6]$ but blue in $[3.6] - [4.5]$ while dusty galaxies undergoing active star formation are red in both colors. From these color behaviors, we determined the color criteria of the BBGs at $z \gtrsim 5$ as follows: 
\begin{eqnarray*}
K - [3.6] > 1.3 \\
K - [3.6] > 2.4 ([3.6] - [4.5]) + 0.6
\end{eqnarray*}
The diagonal boundary in the $K - [3.6]$ versus $[3.6] - [4.5]$ two-color diagram is set parallel to the reddening vector at $z = 5.3$ (black arrows in Figure~\ref{fig4}) so that this color classification is effective even if evolved galaxies have moderate amount of dust. Extremely metal-poor galaxies without dust do not satisfy this BBG criteria even if they are enough old compared with their star-formation timescale. Eight candidate BBGs are selected, which are shown in the two-color and color-magnitude diagrams of Figure~\ref{fig5}. 

The BBG candidates have large uncertainties in the $K - [3.6]$ color measurements because the $K$-band imaging is shallower than the SEDS 3.6$\mu$m and 4.5$\mu$m imaging. In the color-magnitude diagram of Figure~\ref{fig5} (right), we show the peaks of $K - [3.6]$ color distribution for all objects and the associated 4$\sigma$ uncertainties as a function of 3.6$\mu$m magnitudes. The 4$\sigma$ color upper limits for the objects with the peak colors are below $1.3$ and the number of objects is less than 10000 mag$^{-1}$ down to $[3.6] = 24.5$, which implies that typical foreground or background objects are less likely to satisfy our color criteria. 

We also checked stellar contamination in our sample. First, red giant stars in the Milky Way (MW) were ruled out because of their bright absolute magnitudes ranging from $-5.3$ to $-1.5$ in the 3.6$\mu$m \citep{Wainscoat92,Johnson66}, which correspond to the apparent magnitudes ($[3.6] < 18$) brighter than our BBGs out to 100\,kpc distance. Second, we investigated the infrared colors of the brown dwarfs, which might have red $K - [3.6]$ colors due to the cool temperature and methane absorption \citep{Leggett10}, using the catalog of \citet{DupuyLiu12}. We found that the T7.5 or later type dwarf stars have $K - [3.6]$ colors redder than $1.3$ at the absolute magnitude range of $[3.6] = 17.5 - 19.5$, but their $[3.6] - [4.5]$ colors are extremely red ($[3.6] - [4.5] > 1$). We conclude that the BBG candidates are not MW stars but galaxies.

\subsection{\it Multi-wavelength constraint}\label{sec:StrictSelection}
\indent

For the eight BBG candidates, we combined multi-band photometry from $B$ to 24$\mu$m to make SEDs. We also checked the Herschel band-merged catalog \citep{Wang14}, but none of the eight BBGs has counterpart down to 8.8\,mJy at 250$\mu$m ($3\sigma$). 

The eight BBG candidates may still include some contaminants. Several authors suggested that the strong emission lines from young star-forming galaxies potentially affect the broad-band photometry for high redshift galaxies \citep{SchaererdeBarros09,SchaererdeBarros10,Ono10,Inoue11,Stark13}. Especially, strong [OIII] lines from $z \sim 4.5$ galaxies or H$\alpha$ lines from $z \sim 6$ galaxies can make their $K - [3.6]$ colors red. Also, some unforeseen effects caused by completely blending foreground objects might change the colors so that the objects satisfy the BBG color criteria. It is important to identify the most likely candidates of evolved galaxies using the multi-photometry data. For the eight color selected BBGs, we required them to be undetected in the $B$, $V$, $R$, $i'$, $z'$, and 24$\mu$m bands ($> 2\sigma$). The observed $B$, $V$, and $R$ bands correspond to shorter wavelengths than the Lyman break in the rest frame for $z \gtrsim 5$ galaxies. The $i'$ and $z'$ non-detection requirement was adopted to reject the strongest nebular line emitters that have strong UV continuum or Ly$\alpha$ line in those bands. The 24$\mu$m non-detection requirement was adopted to reject local ultra-dusty galaxies ($A_V > 10$) and Type-2 AGNs. After excluding objects detected in the $B$, $V$, $R$, $i'$, $z'$, and 24$\mu$m bands, we obtained only three BBGs as the most likely evolved galaxy candidates. The observed properties of these three most likely BBGs are shown in Table~\ref{tb4}. 



\begin{longtable}{lccccc}
\caption{Observed properties of the most likely BBGs\label{tb4}}
\hline
{Name} & {RA} & {Dec} & {$K$} & {$[3.6]$} & {$[4.5]$} \\
\hline
\endhead
\hline
\hline
\endlastfoot
SEDS\_UDS\_BBG-23 & \timeform{02h16m38s.18} & \timeform{-05D13'52''.1} & 25.24$\pm$0.43 & 23.75$\pm$0.13 & 23.73$\pm$0.13 \\
SEDS\_UDS\_BBG-33 & \timeform{02h17m20s.23} & \timeform{-05D09'07''.5} & 25.45$\pm$0.52 & 23.98$\pm$0.14 & 23.66$\pm$0.13 \\
SEDS\_UDS\_BBG-34 & \timeform{02h17m15s.66} & \timeform{-04D57'59''.8} & $>$ 25.5 ($2\sigma$) & 24.03$\pm$0.15 & 23.79$\pm$0.13 \\
\end{longtable}

\section{Stacking analysis}

\subsection{\it Image stacking} \label{sec:imstack}

\begin{figure*}
\begin{center}
\includegraphics[width=1.0\linewidth, angle=0]{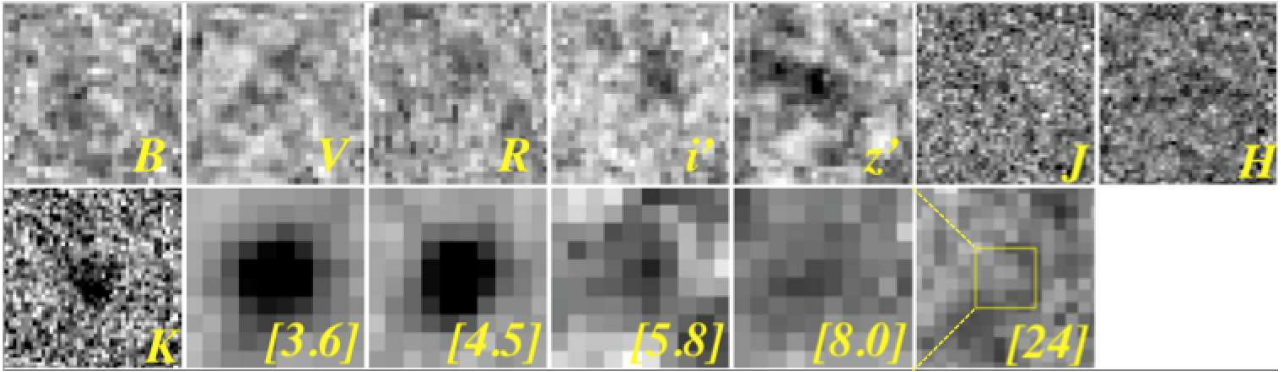}
\end{center}
\caption{Stacked mosaics of the three most likely BBGs. Each panel is $6''$ on a side for the $B - 8.0\mu$m images, and $18''$ for the $24\mu$m image. \label{fig6}}
\end{figure*}

\indent

We made the median stacked images in the $B$, $V$, $R$, $i'$, $z'$, $J$, $H$, $K$, 3.6$\mu$m, 4.5$\mu$m, 5.8$\mu$m, 8.0$\mu$m, and $24 \mu$m for the three most likely BBGs to reduce the background noise fluctuations and constrain their average SEDs. The stacking of only three objects may cause accidental superposition of very faint foreground objects in the photometric aperture. In order to check such superposition effect, we made the 800 different stacked images for each band by rotating the individual images randomly. For these 800 stacked images we performed photometry to evaluate the average magnitude and standard deviation, and finally selected one which yields the average magnitude. The standard deviations in magnitude, which are caused by image rotations, are combined with those of background noise fluctuations to estimate the photometric errors. The stacked images typically become $0.5 \sim 1.0$ mag deeper than the individual frames. 

The stacked images are shown in Figure \ref{fig6}. We have no detection in the $B$, $V$, $R$, $i'$, $J$, $H$, 5.8$\mu$m, and 8.0$\mu$m, but these non-detections yield strong constraint in the SED fitting below. The marginal detection in the stacked $z'$-band image is consistent with the picture that they are galaxies with the relatively faint UV continuum and weak Lyman breaks at $z \gtrsim 5$, rather than actively star-forming galaxies with the strong UV continuum at the same redshift. 

\subsection{\it SED fitting} \label{sec:SEDfit}

\begin{table*}
\caption{Spectral templates and parameter range on SED fitting\label{tb3}}
\begin{center}
\begin{tabular}{ccccccc}
\hline
\hline
Model\footnotemark[$*$] & SFH & Metallicity & Age [Gyr] & $z$\footnotemark[$\dagger$] & A$_V$\footnotemark[$\ddagger$] \\
\hline
BC03 & Exp decline & 0.0001, 0.004, and 0.02 & 0 - t$(z)$\footnotemark[$\S$] & 0 - 7 & 0-4\\
 & ($\tau$ = 0.01, 0.1, 1, and 10 Gyr), &  &  &  & \\
BC03+Nebular line & Constant-SFR & 0.0001, 0.004, and 0.02 & 0 - t$(z)$\footnotemark[$\S$] & 0 - 7 & 0-2 \\
\hline
\end{tabular}
\end{center}
\begin{tabnote}
\footnotemark[$*$] {Spectral template model. ``BC03'' represents stellar population synthesis model of \citet{BruzualCharlot03}. ``BC03+Nebular line'' is the composite model of BC03 stellar continuum and nebular emission lines (see the text or \cite{Salmon??} for detail).}\\
\footnotemark[$\dagger$] {Redshift step is $\Delta z = 0.01$.}\\
\footnotemark[$\ddagger$] {Dust extinction step is $\Delta$A$_V = 0.1$.}\\
\footnotemark[$\S$] {t$(z)$ is the age of the universe at given redshifts.}\\
\end{tabnote}
\end{table*}

\begin{figure*}
\begin{center}
\includegraphics[width=1.0\linewidth, angle=0]{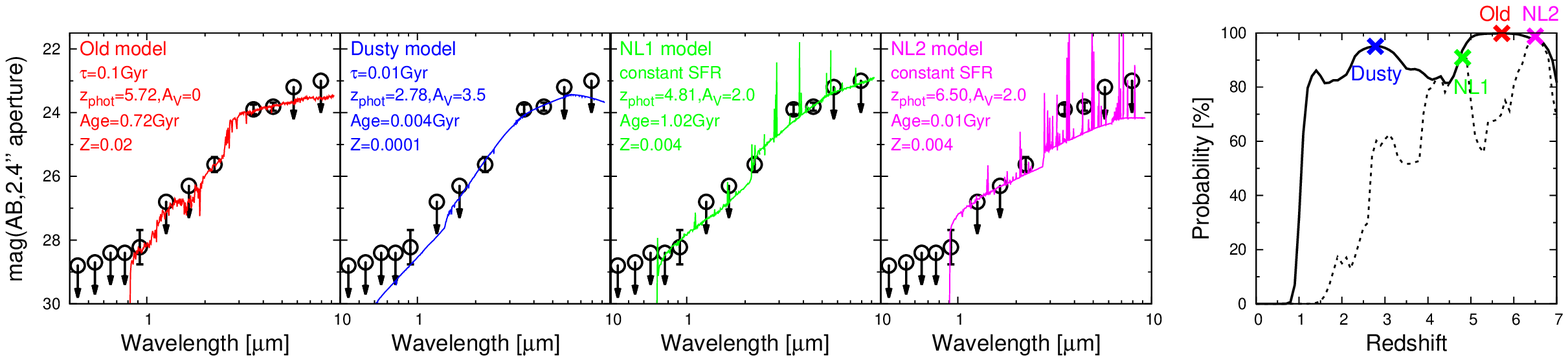}
\end{center}
\caption{Stacked SED of the three most likely BBGs (black circles) and the acceptable-fitting templates from the BC03 or NL spectral templates. Black arrows mean that the flux points are the 2$\sigma$ upper limit. In the far right-hand panel, we also show the probability distribution as a function of redshift. The probability distribution for only the NL templates are shown by the dashed line. \label{fig7}}
\end{figure*}

\indent

For the stacked images, we re-performed the photometry in the same manner as that for individual objects (subsection \ref{sec:multi-phot}). After obtaining the multi-band aperture photometry, we performed SED fitting in the wavelength range between 0.4$\mu$m and 8.0$\mu$m to study the averaged properties of the BBGs at $z \gtrsim 5$ (redshift, SFH, age, stellar mass, and so on). We used the public SED fitting code Hyperz \citep{Bolzonella00}. By setting $\rm{flux} = 0$ and $\Delta\rm{(flux)} = 2\sigma$ for the flux fainter than 2$\sigma$, we kept the number of the measurements larger than that of the fitting parameters and avoided the obvious overfitting. 

Spectral templates consist of the BC03 models with exponentially declining star formation rate (SFR; $\tau$ = 0.01, 0.1, 1, and 10\,Gyr) and the three different metallicity values ($Z = $0.0001, 0.004, and 0.02), which are the same as the templates used in Figure~\ref{fig4}. For them, redshift is changed from 0 to 7 and dust extinction $A_V$ is changed from 0 to 4. Model ages are forced to be less than the universe age at given redshift. We also used spectral templates with strong nebular emission lines (NL templates). In the NL templates, the BC03 model templates with constant SFR as the stellar continuum component. Our calculation of the nebular emission line flux is the same as that in \citet{Salmon??}. The number of ionizing photons was estimated from the SFH, age, and metallicity of each given model, from which we obtained H$\beta$ line flux assuming the case B recombination and an escape fraction of ionizing photons $f_{esc} = 0$. The nebular emission line flux is calculated from their H$\beta$ line flux and the line ratio table of \citet{Inoue11} who used Cloudy 08.00 \citep{Ferland98} to derive nebular spectra as a function of metallicity. We attenuated both nebular line flux and stellar continuum with the same dust extinction, A$_V$, ranging from 0 to 2. Absorption of ionizing photons (LyC absorption) by dust in the ionized nebulae was neglected in this study. Let us take the cases where nebular emission lines contribute broad-band flux at the maximum ($f_{esc} = 0$ and no LyC absorption) for the aim of this study. Our settings for the SED fitting are summarized in Table~\ref{tb3}. We adopted the Calzetti dust extinction law and the Chabrier IMF for all templates. 

The measured SED with the acceptable-fitting templates and the fitting probability distribution which Hyperz software calculated are shown in Figure~\ref{fig7}. The best-fitting template is the old stellar population model at $z = 5.7$ with star-formation timescale $\tau = 0.1$\,Gyr, age $ \sim 0.7$\,Gyr, SFR $= 0.9$\,M$_\odot$\,yr$^{-1}$, no dust, and no nebular line. However, the dusty model at $z = 2.8$ with A$_V = 3.5$ is also acceptable. We further checked this dusty solution using the $24 \mu$m flux which is not used in the SED fitting. Dusty galaxies with A$_V = 3.5$ should have the significant dust emissions at longer wavelength and may be detectable in the 24$\mu$m image. We estimated the infrared luminosity (L$_{\rm IR}$) of the dusty template at $z = 2.8$ assuming that the luminosity attenuated by dust at 0.01$\mu$m $< \lambda <$ 2.2$\mu$m is re-radiated at 5$\mu$m $< \lambda <$ 1000$\mu$m (L$_{\rm IR} = $L$_{\rm abs}$), which results in L$_{\rm IR} = 2.3 \times 10^{12}$\,L$_{\odot}$. We redshifted the model SED of Luminous Infrared Galaxy (LIRG; \cite{Rieke09}) with the same L$_{\rm IR}$ to predict the 24$\mu$m magnitude of the dusty template, and obtained $[24] = 20.5$\,mag. This is inconsistent with the measured flux, $[24] > 21.1$ ($2 \sigma$), which suggests that the dusty solution is unlikely. 

We show the probability distribution for only the NL templates in Figure~\ref{fig7} (dashed line in the right-hand panel). The strong nebular line emitter templates at $z = 4.8$ and $6.5$ are also acceptable. While the strong nebular line emitter model cannot be rejected, we do not consider that all of the BBGs are nebular line emitters. The nebular line emitter solutions for the stacked SED have high stellar mass (M$_{*} > 10^{10}$\,M$_{\odot}$) and high dust extinction (A$_V \sim 2$). Extremely strong nebular line emitters with M$_{*} > 10^{10}$\,M$_{\odot}$ and A$_V \sim 2$ are very rare at $z \sim 2$ \citep{van-der-Wel11,Maseda14} and negligible especially in the photometrically selected passive galaxy sample at $z \sim 4$ \citep{Straatman14}. 

From these above, we conclude that our BBG sample is expected to contain evolved galaxies at $z \gtrsim 5$. Their physical properties are expected from the best-fitting template for the stacked SED: age $= 0.7 \pm 0.4$\,Gyr, M$_{*} = (7.5 \pm 1.5) \times 10^{10}$\,M$_{\odot}$, SFR $= 0.9 \pm 0.2$\,M$_\odot$\,yr$^{-1}$ and A$_V = 0^{+1}_{-0}$ at $z = 5.7 \pm 0.6$, where the uncertainties correspond to the 1$\sigma$ confidence interval ($\Delta \chi^2 < 1$) for each quantity. They are massive, but are consistent with the stellar mass of the analogs identified at intermediate redshifts ($z = 2 -4$: \cite{Muzzin13,Spitler14}).


\section{Discussion}\label{discussion}
\indent

In the following discussion about the physical properties of the evolved galaxies in the early universe, we roughly assume that all of the three BBG candidates have a stellar masses of M$_* > 5 \times 10^{10}$\,M$_\odot$ and ages older than 0.3\,Gyr at the redshift $5 < z < 7$. This assumption is well supported by the stacking analysis, while we cannot rule out the possibility that some of them are massive, dusty, but are extremely strong nebular line emitters (subsection \ref{sec:SEDfit}).

\subsection{\it Implication about the stellar mass density of evolved galaxies at $z > 5$}

\begin{figure}
\begin{center}
\includegraphics[width=0.9\linewidth, angle=0]{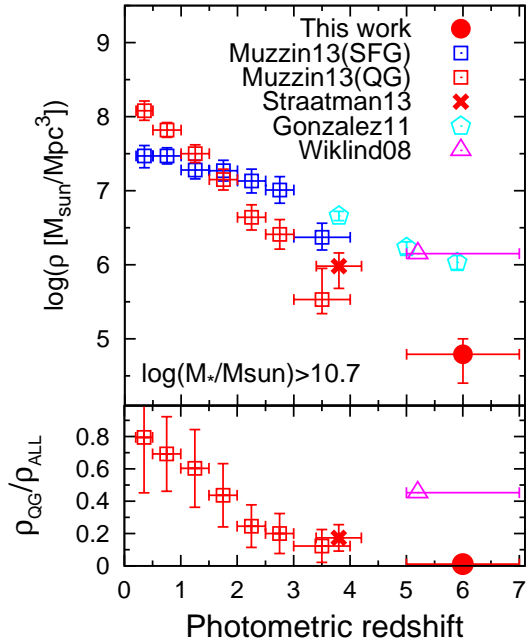}
\end{center}
\caption{(Top panel) Evolution of the stellar mass density in the universe between $z = 0 - 7$. Our measurements for evolved galaxies at $z \sim 6$ are shown by red filled circles. Other symbols correspond to the measurements from previous studies: red squares for the quiescent galaxies at $z = 0 - 4$ \citep{Muzzin13}, a red cross for the quiescent galaxies at $z \sim 4$ \citep{Straatman14}, a magenta triangle for the evolved galaxies at $z \sim 5.2$ \citep{Wiklind08}, blue squares for the star-forming galaxies at $z = 0 - 4$ \citep{Muzzin13}, and cyan pentagons for the LBGs at $z = 4 - 6$ \citep{Gonzalez11}. All of them are corrected so that only galaxies with $\log($M$_{*}/$M$_{\odot}) \geq 10.7$ contribute them. (Bottom panel) Quiescent fraction in stellar mass densities. Symbols are the same as those in the top panel. The sum of quiescent and star-forming galaxies are used as the total stellar mass densities, where we used the LBGs of \citet{Gonzalez11} as star-forming galaxies for the quiescent galaxies of \citet{Straatman14}, \citet{Wiklind08}, and this work.  \label{fig8}}
\end{figure}

\indent

We estimated the number and stellar mass density of evolved galaxies at $z \gtrsim 5$. The survey volume was calculated from the SEDS area (0.34\,deg$^2$) and the redshift range ($z = 5 - 7$). The survey incompleteness was also corrected by using the completeness of our source extraction at the 3.6$\mu$m magnitude range of the BBGs ($[3.6] = 23.5 - 24.0$), $56\pm1$\,\% (see subsection~\ref{sec:multi-phot} and Figure~\ref{fig3}). The estimated number and stellar mass density of evolved galaxies at $z = 5 - 7$ are $(8 \pm 5) \times 10^{-7}$\,Mpc$^{-3}$ and $(6 \pm 4) \times 10^{4}$\,M$_\odot$\,Mpc$^{-3}$, respectively. These densities may be underestimated, because we might have missed some evolved galaxies as a result of photometric uncertainties. We found that about 30\,\% of objects with an intrinsic color of $K - [3.6] > 1.3$ are not observed as BBGs after considering the random photometric errors which are extracted from Gaussian probability distributions. This effect, however, does not change the following discussions very much.

We compared this stellar mass density of evolved galaxies at $z = 5 - 7$ with those at $z < 4$ of \citet{Muzzin13} and \citet{Straatman14}. \citet{Muzzin13} used  the COSMOS/UltraVISTA survey data to select both star-forming and quiescent galaxies up to $z = 4$ and to construct their stellar mass functions. Their classification of quiescent and star-forming galaxies using rest $U - V$ and $V - J$ colors is qualitatively similar with our BBG selection, in the sense that both selections aim to isolate galaxies with red color by Balmer breaks and blue color in longer wavelength than Balmer breaks. We estimated the stellar mass densities at $z = 0 - 4$ by integrating their stellar mass functions down to $\log($M$_{*}/$M$_{\odot}) = 10.7$, which equals the stellar mass lower limit assumed for our BBG sample. \citet{Straatman14} used the FourStar Galaxy Evolution Survey (ZFOURGE) data to construct quiescent galaxy sample down to $\log($M$_{*}/$M$_{\odot}) = 10.6$ at $z \sim 4$, where their selection method was the same as that in \citet{Muzzin13} (rest-$UVJ$ color selection). For the 15 quiescent galaxies at $z \sim 4$, we adopted our stellar mass limit to calculate the stellar mass density. We show these stellar mass densities of quiescent galaxies with $\log($M$_{*}/$M$_{\odot}) \geq 10.7$ in Figure \ref{fig8}. The stellar mass density at $z \sim 6$ revealed in this paper is much lower than that at $z = 3 - 4$ and consistent with the trend that the stellar mass density of quiescent galaxies decrease continuously with increasing redshift.

We also compared the stellar mass density of evolved galaxies at $z \gtrsim 5$ with those of \citet{Wiklind08}. The color selection of \citet{Wiklind08} for BBGs at $z \gtrsim 5$ is similar to our BBG selection in which case they require red $K - [3.6]$ colors, but different from our because they remove the lower redshift contaminations by using the blueward colors ($J - K$ and $H - [3.6]$) instead of $[3.6] - [4.5]$. They identified four candidates for evolved galaxies at $z = 5 - 7$ without the 24$\mu$m detection in the 0.04\,deg$^2$ area of the GOODS-South field which have the physical properties similar to ours: $\log {\rm M}_* = 10.7 - 11.7$ and ${\rm age} = 0.2 - 1$\,Gyr. The stellar mass density after correcting the detection incompleteness is $1.4 \times 10^6$\,M$_\odot$\,Mpc$^{-3}$, which is much more than our measurement (Figure~\ref{fig8}). We cannot give a conclusive explanation for this inconsistency, although it is partially due to the fact that they did not use the nebular line emitter templates.

In Figure \ref{fig8}, the stellar mass densities of star-forming galaxies are also plotted; they were estimated by integrating the stellar mass functions of star-forming galaxies at $z = 0 - 4$ \citep{Muzzin13} and Lyman Break Galaxies (LBGs) at $z = 4 - 6$ \citep{Gonzalez11} down to $\log($M$_{*}/$M$_{\odot}) = 10.7$. From these, we estimated the quiescent fraction, which is defined as the fraction of stellar mass density of quiescent galaxies compared with the total stellar mass density. Here we assume the sum of stellar mass density of quiescent galaxies and star-forming galaxies (LBG at $z = 4 - 7$) as the total stellar mass density. The quiescent fraction at $z \sim 6$ is only $0.01\pm0.0009$, which is much lower than $\sim 0.15$ at $z \sim 4$. This suggests that evolved galaxies do not dominate the massive end of stellar mass functions at $z > 5$. Such rare but non-zero evolved galaxies are naturally considered to be embedded in prominent massive halos, because both theoretical works (e.g., \cite{DeLucia06}) and observations \citep{Tanaka10,Kubo13} suggested that mass assembly and quenching of galaxies preferentially proceeded in the proto-cluster regions. It is worthwhile to investigate the surrounding environments of the evolved galaxy candidates individually in future.

\subsection{\it Possible progenitor of evolved galaxies at $z > 5$}

\begin{figure}[]
\begin{center}
\includegraphics[width=1.0\linewidth, angle=0]{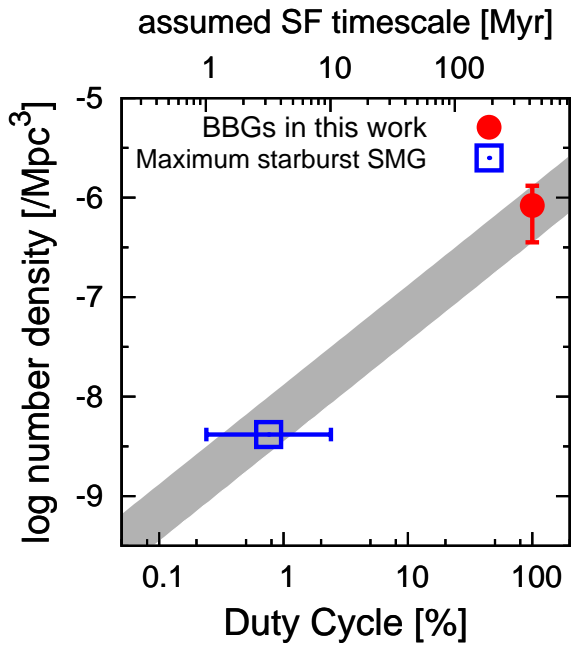}
\end{center}
\caption{The number densities of the evolved galaxy candidates (red filled circles) and ``maximum starburst'' SMGs (blue open squares) at $z = 5 - 7$ as a function of duty cycle. Duty cycle for ``maximum starburst'' SMGs are estimated from their star-forming timescale. The expected region for progenitor of evolved galaxies is shown by grey shading. \label{fig9}}
\end{figure}

\indent

Finally, we discuss the progenitor of the evolved galaxies at $z \gtrsim 5$. Our finding of evolved galaxies at $z > 5$ strongly suggests intense starburst in these systems at much larger redshift. Submilimeter-selected galaxies (SMGs) are considered as possible progenitors, because their dust-obscured star formation is much more active than that of UV-selected star-forming galaxies which leads to rapid consumption of gas and subsequent quenching \citep{Toft14}. 
 
Although very few SMGs with spectroscopic redshifts are available above $z = 5$, the object HFLS3 \citep{Riechers13} may be a good example of a progenitor of the evolved galaxies at $z \gtrsim 5$. HFLS3 was identified by ultra red colors in the Herschel SPIRE wavelength range ($\lambda = 250 - 500 \mu$m), and its redshift was determined to be $z = 6.3369 \pm 0.0009$ based on several emission lines \citep{Riechers13}. They found that HFL3 has an extremely high SFR ($2,900$\,M$_{\odot}$\,yr$^{-1}$), a large amount of gas ($1.0 \times 10^{11}$\,M$_{\odot}$), and large stellar mass ($3.7 \times 10^{10}$\,M$_{\odot}$). These quantities are not changed very much even if the lensing magnification by the two foreground galaxies is taken into account \citep{Cooray14}. Although the gas depletion timescale of HFLS3 is 36\,Myr, star formation may be suppressed on a shorter timescale ($\sim 1 - 10$\,Myr) because of supernova (and AGN) feedback. If HFLS3 continue to produce stars with SFR$ = 2,900$\,M$_{\odot}$\,yr$^{-1}$ for 10\,Myr and experiences subsequent passive evolution, it should become BBG with M$_{*} = 6.6 \times 10^{10}$\,M$_{\odot}$ and age $\sim 0.3$\,Gyr at $z \sim 5$, which is consistent with stellar mass and age of our evolved galaxy candidates within the $1 \sigma$ uncertainties. This suggests that the most active starburst (or ``maximum starburst'') SMGs can be progenitors of evolved galaxies at $z \gtrsim 5$. 

We also checked their abundances. The expected number density of progenitors depends on their duty cycle. We here define duty cycle as the ratio of the time duration for which galaxies can be detected to the time interval that corresponds to the searched redshift range. Once galaxies are quenched, they can be observed as evolved galaxies at all times, which means that the duty cycle of evolved galaxies is 100\,\%. If a progenitor has a duty cycle of 100\,\%, the expected number density should be same as the observed number density of evolved galaxies, $8 \times 10^{-7}$\,Mpc$^{-3}$. But the duty cycle of a star-forming progenitor may be small, because the time duration for which it was in active star-forming phase is expected to be short. The number density of such a progenitor should be $8 \times 10^{-7}$\,Mpc$^{-3} per $duty-cycle. For ``maximum starburst'' SMGs such as HFLS3, we adopt the following simple assumptions. Intrinsic redshift distribution is flat at $z = 5 - 7$. Completeness of the Herschel observation of \citet{Riechers13}, which is a part of HerMES \citep{Oliver12} in the First Look Survey (FLS) field, Locman Hole, and GOODS-North fields, for galaxies analog to HFLS3 at $z = 5 - 7$ is 0.6. This completeness is consistent with that estimated in \citet{Wang14} for objects with flux of 30\,mJy at 500$\mu$m (flux limit of \cite{Riechers13}). Observed time interval corresponding to $z = 5 - 7$ is 0.42\,Gyr, and star-formation timescale of HFLS3 is $1 - 10$\,Myr. These assumptions lead to a spatial number density of $4.2 \times 10^{-9}$\,Mpc$^{-3}$ and duty cycle of $0.24 - 2.4$\,\% for ``maximum starburst'' SMGs at $z = 5 - 7$. The number densities of ``maximum starburst'' SMGs at $z = 5 - 7$ and expected progenitor as a function of duty cycle are shown in Figure \ref{fig9}. The number density of ``maximum starburst'' SMGs is consistent with that of expected progenitors. \citet{Riechers13} reported on the four analogs of HFLS3, which have similar red Herschel SPIRE colors. They may explain the small difference between the number density of ``maximum starburst'' SMGs and expected progenitors, while there are large uncertainties caused by the simple assumptions. The picture that star-formation in the progenitor of evolved galaxies at $z \gtrsim 5$ rapidly proceeds surrounded with dust is a natural extension of the picture for lower-redshift, quiescent galaxies \citep{Toft14,Straatman14}.

\section{Summary}
\indent

In this paper, we have investigated the nature of candidate evolved galaxies at $z \gtrsim 5$. We identified eight galaxies with $K - [3.6] > 1.3$ and $K - [3.6] > 2.4 ([3.6] - [4.5]) + 0.6$ down to $[3.6] \sim 24$\,mag from the SEDS UDS 0.34\,deg$^2$ area. Among them, we further selected the three most likely BBGs at $z \gtrsim 5$ by requiring non-detection in the $B$, $V$, $R$, $i'$, $z'$, and 24$\mu$m bands. The best-fitting spectral template for the stacked SED is the old stellar population model with M$_{*} = (7.5 \pm 1.5) \times 10^{10}$\,M$_{\odot}$, SFR $= 0.9 \pm 0.2$\,M$_\odot$\,yr$^{-1}$, and age $= 0.7 \pm 0.4$\,Gyr at $z = 5.7 \pm 0.6$. We estimated the number and stellar mass densities of evolved galaxies at $z \gtrsim 5$ down to $\log($M$_{*} / $M$_{\odot}) = 10.7$ assuming all of the three most likely BBGs are really such objects, which results in $(8 \pm 5) \times 10^{-7}$\,Mpc$^{-3}$ and $(6 \pm 4) \times 10^{4}$\,M$_\odot$\,Mpc$^{-3}$, respectively. This stellar mass density of evolved galaxies at $z = 5 - 7$ is consistent with the declining trend previously observed within a range of $0 < z < 4$. A very simple abundance-matching test implies that the evolved galaxies identified in this paper seem to be the descendants of ``maximum starburst'' SMGs, such as HFLS3 reported in \citet{Riechers13}. Our finding of evolved galaxy candidates suggests that the first starburst and subsequent quenching process have been completed by $z \sim 6$. 

The BBGs are too red and too faint for us to further constrain their detailed properties against the photometric uncertainties. These galaxies will, however, be fully revealed with the next-generation infrared instruments on the future {\it Wide-field Imaging Surveyor for High-redshift} ({\it WISH}: \cite{Yamada12}) or {\it James Webb Space Telescope} ({\it JWST}: \cite{Gardner06}).

\bigskip

We thank Brett Salmon and Steven Finkelstein for the code they offered which calculates nebular emission line flux. We also appreciate Hisanori Furusawa for kindly offering the new Subaru $z'$-band imaging. This research was conducted as a part of K. M.'s phD thesis in Tohoku University. K. M. is grateful to Takashi Ichikawa, Masafumi Noguchi, and Masayuki Akiyama for valuable comments and suggestions. As this research started during K. M. stayed at the Harvard-Smithsonian Astrophysical Observatory, K. M. appreciates the SAO predoctoral program. This research is financially supported by Japan Society for the Promotion of Science through Brain Circulation Program (R2301) and Grant-in-aid for Scientific Research (13J02968 and 26400217).

\end{document}